\documentclass[%
 reprint,
superscriptaddress,
 amsmath,amssymb,
 aps,
prb,
]{revtex4-1}

\usepackage{color}
\usepackage{graphicx}
\usepackage{breqn}
\usepackage{dcolumn}
\usepackage{bm}

\usepackage{xspace}

\newcommand{\ket}[1]{{| #1 \rangle}}
\newcommand{\bra}[1]{{\langle  #1 |}}

\begin{document}

\preprint{APS/123-QED}

\title{Generalized slave-particle method for extended Hubbard models}

\author{Alexandru B. Georgescu}
\affiliation{Department of Physics}
\affiliation{Center of Research on Interface Structure and Phenomena}
\author{Sohrab Ismail-Beigi}
\affiliation{Department of Applied Physics}
\affiliation{Department of Physics}
\affiliation{Center of Research on Interface Structure and Phenomena}
\affiliation{Department of  Mechanical Engineering and Materials Science\\Yale University$,$ New Haven$,$ CT 06520$,$ USA}




\date{\today}

\begin{abstract}

We introduce a set of generalized slave-particle models for extended Hubbard models that treat localized electronic correlations using slave-boson decompositions.  Our models automatically include two slave-particle methods of recent interest, the slave-rotor and slave-spin methods, as well as a ladder of new intermediate models where one can choose which of the electronic degrees of freedom (e.g., spin or orbital labels) are treated as correlated degrees of freedom by the slave bosons.   In addition, our method removes the aberrant behavior of the slave-rotor model, where it systematically overestimates the importance of electronic correlation effects for weak interaction strength, by removing the contribution of unphysical states from the bosonic Hilbert space.  The flexibility of our formalism permits one to separate and isolate the effect of correlations on the key degrees of freedom. 
%
\end{abstract}

\maketitle


\section{Introduction}

One of the long-standing areas of interest in condensed matter physics, particularly that of complex oxides, is that of the Mott metal-insulator transition\cite{Imada1998}.  Generically, within a Hubbard model framework, as the strength of localized electronic repulsions is increased, the electrons prefer to be localized on atomic sites and inter-site hopping is suppressed, and at a critical interaction strength the system becomes an insulator.  An illustration of the rich behavior that can occur in such systems is the Orbital Selective Mott Transition (OSMT) whereby only a subset of localized orbitals become insulating (localized) while the remainder have metallic (extended) bands.  An example is provided by quasi-two-dimensional Mott transition in the Ca$_{2-x}$Sr$_x$RuO$_4$ family, where the Mott metal-insulator transition and its magnetic properties \cite{Nakatsuji2000} at the critical doping $x=0.5$ show a coexistence between a magnetic susceptibility that shows a Curie form for $S=1/2$ and a metallic state. Anisimov {\it et al.} \cite{Anisimov2002} have used DFT+DMFT to explain this situation in terms of an OSMT in which one Ru $4d$ orbital is localized, while the other continues to present metallic behavior.

The present day workhorse for {\it ab initio} materials modeling and prediction, Density Functional Theory (DFT), is fundamentally based on band theory and is unable to describe such transitions (without symmetry breaking of the electronic degrees of freedom: e.g., spin or orbital polarization).  To this end, Hubbard model based methods such as Dynamical Mean Field Theory (DMFT) and DFT+DMFT\cite{Georges1996,Kotliar2004} have been developed to include localized correlation effects in electronic structure calculations.  However, DMFT-based methods are computationally expensive and typical present day calculations on real materials are generally restricted to treating a few correlated sites.  Therefore, it is of significant interest to have computationally inexpensive, but necessarily more approximate, methods that include correlations and can permit one to rapidly explore the qualitative effects of electronic correlations.

One set of such approximate methods that have been of recent interest are slave-particle methods.  Slave-boson methods have a long background in condensed matter theory for analytical treatments of correlations typically in the limiting case of infinite correlation strength\cite{Barnes1976,Barnes1977,Coleman1983,Read1983,Read1985,Kotliar1986,Lee2006}.  Kotliar and Ruckenstein\citep{Kotliar1986} used a slave-particle representation to treat Hubbard-like models at finite interaction strength, which found applications in the realm of high-temperature superconductivity\citep{Raczkowski2006}. Further, Kotliar and Ruckenstein's model has been generalized to multi-band models\citep{Fresard1996,Lechermann2007,Bunemann2011} where, e.g., the effects of multiple orbitals, orbital degeneracy, and the Hund's interaction have been studied.\citep{Fresard1996,Lechermann2007}  However, the approach of Kotliar and Ruckenstein, and its various extensions, require a large number of bosonic slave particles: one needs one boson per possible electronic configuration on a correlated site.

For this reason, more economical slave-boson representations have been of significant interest. Florens and Georges\citep{Florens2002, Florens2004} used a single ``rotor'' slave-boson per site that describes the total electron count on each site in a computationally economical manner.  The slave-rotor method has  successfully predicted a number of electronic phases of nickelate heterostructures\citep{Lau2013a} which was a distinct improvement over previous studies.  However, a rotor-like description is not orbitally selective as it can only describe the total electron count on a site and not its partitioning among inequivalent orbitals on that site.  An alternative slave-particle approach is to treat each localized electronic state (i.e., a unique combination of spin and orbital indices) with a slave boson: this ``slave-spin'' approach automatically handles orbital symmetry breaking and can predict OSMTs\citep{DeMedici2005,Hassan2010}.  Recently, it has been applied to predict key physical characteristics in iron superconductors \cite{DeMedici2014}. 

In this work, we introduce a generalized framework for slave-particle descriptions.  This produces a ladder of correlated models, and the slave-rotor and slave-spin are automatically included as two specific cases.  Our approach does not require any physical analogies to create the slave bosons (e.g., a quantum rotor or angular variable to motivate the slave-rotor or a pseudo-spin to motive the slave-spin) and works directly in the occupation number representation.  In our approach, one can choose which degrees of freedom are treated as correlated degrees of freedom (e.g., total electron count on a site, electron counts in each orbital, electron count in each spin channel, etc.) so that we can isolate the effect of correlations on the separate degrees of freedom in a systematic manner.  Section \ref{sec:basics} presents our general formalism, how it builds upon previous models, as well as gives a few examples of models that can be built within this framework. Section \ref{sec:tests} is devoted to tests of possible models built within this formalism in a mean-field approach at half-filling within a one-band and a two-band model in order to compare our results with those of previous work as well as to better understand the role of the different terms in an extended Hubbard model within our formalism. In Section \ref{sec:conclusion} we conclude this paper and discuss possible new avenues for researchers to use this method and possible developments of it in predicting properties of correlated materials.


\section{The Generalized Slave-Particle Representation}
\label{sec:basics}

In this section we introduce our generalized slave-particle representation. In  appropriate limits, our approach reproduces previous frameworks such as the slave-rotor and slave-spin methods.  One utility of our approach is that it allows us to unite these two, as well as other intermediate models, into a single slave-particle methodology.  A variety of slaves-particle models can be investigated and compared so that one can isolate which specific correlated degrees of freedom are critical for describing a specific physical problem.

\subsection{Extended Hubbard model}

The general correlated-electron Hamiltonian we consider is an extended Hubbard model given by
\begin{multline}
\hat{H}=\sum_i \hat{H}_{int}^i +\sum_{im\sigma}\epsilon_{im\sigma}\hat{d}^\dag_{im\sigma}\hat{d}_{im\sigma}\\
-\sum_{ii'mm'\sigma} t_{imi'm'\sigma}  \hat d^\dag_{im\sigma} \hat d_{i'm'\sigma}\,.
\end{multline}
The index $i$ ranges over the localized sites of the system (usually atomic sites), $m$ ranges over the localized spatial orbitals on each site, $\sigma$ denotes spin, $\hat{H}_{int}^i$ is the local Coulombic interaction for site $i$ detailed further below, $\epsilon_{im\sigma}$ is the onsite energy of the orbital $im\sigma$, and $t_{imi'm'\sigma}$ is the spin-conserving hopping element connecting orbital $im\sigma$ to $i'm'\sigma$.  The $\hat d$ are canonical fermion annihilation operators.  We take the interaction term to have the standard Slater-Kanamori form \citep{Kanamori1963}
\begin{multline}
\hat{H}_{int}^i=\frac{U_i}{2}(\hat{n}_i^2-\hat{n}_i)+\frac{U'_i-U_i}{2}\sum_{m\neq m'}\hat{n}_{im}\hat{n}_{im'}\\
-\frac{J_i}{2}\sum_{\sigma}\sum_{m\neq m'}\hat{n}_{im\sigma}\hat{n}_{im'\sigma}\\
-\frac{J_i}{2}\sum_{\sigma}\sum_{m\neq m'}\left\{
\hat{d}^\dag_{im\sigma}\hat{d}_{im\bar\sigma}\hat{d}^\dag_{im'\bar\sigma}\hat{d}_{im'\sigma}\right.\\
\left. +\hat{d}^\dag_{im\sigma}\hat{d}^\dag_{im\bar\sigma}\hat{d}_{im'\sigma}\hat{d}_{im'\bar\sigma}\right\}\,.
\label{eq:SlaterKanamori}
\end{multline}
The first and second term stem from Coulombic repulsion terms between the same spatial orbital ($U$) and different spatial orbitals ($U'$).  The third term is Hund's exchange between different orbitals of the same spin with strength $J$. The fourth term contains the  intrasite ``spin flip'' and ``pair hopping'' terms.  The index $\bar\sigma$ is the spin opposite to $\sigma$.   The subscripts $i$ on the $U$, $U'$ and $J$ parameters denote the fact that each correlated site can have its own set of parameters; however, to keep indices to a minimum below, we suppress this index.  The various number operators are 
\begin{eqnarray*}
\hat n_{im\sigma} = \hat d_{im\sigma}^\dag\hat d_{im\sigma} \ \ & , & \ \ \hat n_{im} = \sum_\sigma \hat n_{im\sigma} \\
\hat n_{i\sigma} = \sum_m \hat n_{im\sigma} \ \ & , & \ \ \hat n_i =\sum_{m\sigma} \hat n_{im\sigma}\,.
\end{eqnarray*}
For what follows, we keep in mind that due to the fact that $\hat n_{im\sigma}^2=\hat n_{im\sigma}$, the Hund's term in $\hat H^i_{int}$ can be rewritten in an equivalent form to give
\begin{multline*}
\hat{H}_{int}^i=\frac{U_i}{2}(\hat{n}_i^2-\hat{n}_i)+\frac{U'_i-U_i}{2}\sum_{m\neq m'}\hat{n}_{im}\hat{n}_{im'}\\
-\frac{J_i}{2}\sum_{\sigma}(\hat n_{i\sigma}^2-\hat n_{i\sigma})^2\\
-\frac{J_i}{2}\sum_{\sigma}\sum_{m\neq m'}\left\{
\hat{d}^\dag_{im\sigma}\hat{d}_{im\bar\sigma}\hat{d}^\dag_{im'\bar\sigma}\hat{d}_{im'\sigma}\right.\\
\left. +\hat{d}^\dag_{im\sigma}\hat{d}^\dag_{im\bar\sigma}\hat{d}_{im'\sigma}\hat{d}_{im'\bar\sigma}\right\}\,.
\end{multline*}

\subsection{Spinons and slave bosons}

The interacting Hubbard hamiltonian is impossible to solve exactly and even difficult to solve approximately.  Part of the difficulty comes from the fact that we have interacting fermions which have both charge and spin degrees of freedom.  Following well-known ideas in slave-boson approaches\cite{Barnes1976,Barnes1977,Coleman1983,Read1983,Read1985,Kotliar1986,Lee2006}, one separates at each site the fermionic degrees of freedom from the charge degrees of freedom by introducing a bosonic ``slave'' particle on that site.  The boson is spinless and charged, and one also has a remaining neutral fermion with spin termed a spinon.  

With spinons denoted by $\hat f$ operators and slave bosons by $\hat O$ operators, we define
\begin{dmath}
\hat{d}_{im\sigma}=\hat{f}_{im\sigma}\hat{O}_{i\alpha}
\label{eq:slaveopdef}
\end{dmath}
and
\begin{dmath}
\hat{d}_{im\sigma}^{\dagger}=\hat{f}_{im\sigma}^{\dagger}\hat{O}_{i\alpha}^{\dagger}\,.
\label{eq:slaveopdagdef}
\end{dmath}
Requiring the $\hat f$ to be fermionic field operators in turn requires the $\hat O$ operators to obey bosonic commutation relations.  We note that while the $\hat f$ spinon operators are standard fermionic Fock field operators, the bosonic $\hat O$ operators are generic and {\it ad hoc}: there is no assumption or requirement that the $\hat O$ be bosonic operators for a Fock space (and in general they are not of that variety).

The index $\alpha$ is part of our generalized notation that permits us to unify many slave-particle models.  The meaning of $\alpha$ depends on the type of model chosen, as we will show in detail below with a variety of examples.  The index $\alpha$ refers to a subset of the complete set of $m\sigma$ indices that belong to a site $i$.  For example, if we use an O(2) slave-rotor model for the correlated orbitals on an site \citep{Florens2002, Florens2004} where $\hat O = e^{-i\hat \theta}$ and $\theta$ is the phase angle of the O(2) rotor, then $\alpha$ is nil: $\hat O_{i\alpha}=\hat O_i$.  Namely, we have a single slave particle on each site $i$ that tracks the total number of particles on that site.  At the opposite limit, we can have a unique slave boson for each  $m\sigma$ (the ``slave-spin'' method\citep{DeMedici2005, Hassan2010}), so that $\alpha=m\sigma$. 

We work directly in the number representation and introduce a number operator for the slave particles (this is a generalization of the angular momentum operator for the slave-rotor approach or the $S_z^{im\sigma}$ quasi-spin of the slave-spin representation).  The minimum and maximum allowed particle numbers are $N_{min}$ and $N_{max}$ so that 
\begin{equation}
\hat N_{i\alpha} = \mbox{diag }(N_{min},N_{min}+1,\ldots,N_{max}-1,N_{max})\,.
\end{equation}
This operator simply keeps track of the number of slave-particles in each slave mode $i\alpha$, and the minimum and maximum allowed occupancies depends the slave model we choose as discussed below.

Since we have introduced new degrees of freedom and enlarged the Hilbert space of the problem, it is necessary to enforce constraints so that one avoids considering ``unphysical states'' that have no correspondence to those in the original problem.  The original Hilbert space is spanned by kets of the form $\ket{\{n_{im\sigma}\}}$ in the occupancy basis of the $\hat d_{im\sigma}$ operators.  The enlarged Hilbert space is spanned by kets of the product form $\ket{\{n_{im\sigma}\}}_f\ket{\{N_{i\alpha}\}}_s$ where the subscripts label spinon and slave sectors.  Within this enlarged space, there is a subset of ``physical states'' that correspond to the original kets.  The first part of the correspondence is make the fermionic occupancies $\{n_{im\sigma}\}$ of the original electron counts ($\hat d$) and the spinon counts ($\hat f$) identical.  Hence, the real question is which $N_{i\alpha}$ are physically allowed.  

Equations (\ref{eq:slaveopdef}) and (\ref{eq:slaveopdagdef}) mean that the number of spinon and slave particles on each site must track each other because they are annihilated and created at the same time.  Hence, following ideas from prior work,\cite{Florens2002,DeMedici2005} we enforce constraints to ensure the particle numbers track each other.  We enforce the constraint
\begin{dmath}
\sum_{m\sigma \in \alpha}\hat f^\dag_{im\sigma} \hat f_{im\sigma}\ket{\phi} =\hat{N}_{i\alpha}\ket{\phi} 
\label{eq:fmatchNconstraint}
\end{dmath}
on the kets $\ket{\phi}$ in the enlarged space.  Only these kets are physically allowed in the exact description of the system, and they span the physical subspace.  We note that the constraint is on the allowed kets and not on the operators: the operators act in the extended Hilbert space that includes physical and unphysical unphysical states, and the $\hat f$ or $\hat O$ operators acting alone can move us from a physical state to an unphysical state.

Enforcing the constraints of Eq.~(\ref{eq:fmatchNconstraint}) ensures that only physical states that are in one-to-one correspondence to the original states are considered in the extended spinon+slave boson Hilbert space.  (This is the same idea as Ref.~\onlinecite{Florens2002} where the O(2) rotor angular momentum operator $\hat{L}$ has been replaced by $\hat{N}_{i\alpha}$.)  When the constraints are obeyed, only physically allowed occupancies for the bosonic operators are relevant which is the same as setting $N_{min}=0$ and $N_{max}$ to the maximum allowed occupancy in the slave sector on the site.  For example, for a system of 3 spatial orbitals on a state (i.e, three choices of $m$), $N_{max}=6$ if we only wish to count total numbers of electrons using the slave bosons in which case $\alpha$ is nil.  Or, if we want to count up and down spin electrons separately for this site, then $\alpha$ is the same as $\sigma$ and $N_{max}=3$ for each choice of $\alpha$.  However, as explained below, when one does approximate calculations, choosing $N_{min}$ and $N_{max}$ that differ from these values allows for the creation of different types of models and offers some technical advantages.

For completeness and clarity, Appendix~\ref{appendix1} provides explicit examples of the enlarged Hilbert spaces and various choices of $\alpha$ and explains how the action of the $\hat d_{im\sigma}$ and the $\hat f_{im\sigma}\hat O_{i\alpha}$ operators are identical on the physical subspace of states.

\subsection{Slave operators and Hamiltonian}

To reproduce the standard behavior of the annihilation operator where only physical states are allowed in the spectrum of the operator \cite{DeMedici2005,Hassan2010}  (also see the Appendix \ref{appendix1}),
\begin{equation}
\hat{d}_{im\sigma}|n_{im\sigma}\rangle=\sqrt{n_{im\sigma}}|n_{im\sigma}-1\rangle\,.
\label{eq:dlowernumber}
\end{equation}
it must be that
\begin{equation}
\hat{f}_{im\sigma}|n_{im\sigma}\rangle_f=\sqrt{n_{im\sigma}}|n_{im\sigma}-1\rangle_f
\label{eq:flowernumber}
\end{equation}
and
\begin{equation}
\hat{O}_{i\alpha}|N_{i\alpha}\rangle_s = |N_{i\alpha}-1\rangle_s\,.
\label{eq:Olowernumber}
\end{equation}
However, if $n_{im\sigma}=0$, then the action of $\hat f_{im\sigma}$ will destroy the total state $\ket{\phi}$ regardless of what $\hat O_\alpha$ may do, so for this case we have an undetermined situation:
\begin{equation}
\hat{O}_{i\alpha}|N_{i\alpha}=0\rangle_s=\mbox{undetermined}\,.
\end{equation}
Following the same logic for the creation operators yields
\begin{equation}
\hat{O}_{i\alpha}^{\dagger}|N_{i\alpha}\rangle_s=|N_{i\alpha}+1\rangle_s
\end{equation}
until we reach the ceiling $N_{i\alpha}=N_{max}$ when we have a similar indeterminacy
\[
\hat{O}_{i\alpha}^{\dagger}|N_{max}\rangle_s=\mbox{undetermined}\,.
\]

Putting this all together, the slave boson operator $\hat O_{i\alpha}$  in the number basis must have the form
\begin{equation}
\hat{O}_{i\alpha}= \left( \begin{array}{cccccc}
0 & 1 & 0 & \ldots & 0 & 0\\
0 & 0 & 1 & \ldots & 0 & 0\\
\vdots & \vdots & \vdots & \ddots & \vdots & \vdots \\
0 & 0 & 0 & \ldots & 1 & 0\\
0 & 0 & 0 & \ldots & 0 & 1\\
C_{i\alpha} & 0 & 0 & \ldots & 0 & 0\end{array} \right) 
\end{equation}
where $C_{i\alpha}$ is at this point an undetermined constant that we are free to choose.  Below, we will use this freedom to ensure that we reproduce a desired non-interacting band structure at zero interaction strength (when $\hat H_{int}^i=0$).

Substituting the spinon and slave operators into the original extended Hubbard Hamiltonian gives the following form, which for the moment we specialize to the symmetric $U'=U, J=0$ case to keep the logic simple (the more general cases are enumerated  further below):
\begin{multline*}
\hat{H}=\frac{U}{2}\sum_i\left((\sum_{\alpha}\hat{N}_{i\alpha})^2-\sum_{\alpha}\hat{N}_{i\alpha}\right)\\
+\sum_{im\sigma}\epsilon_{im\sigma}\hat{f}^\dag_{im\sigma}\hat{f}_{im\sigma}\\
-\sum_{ii'mm'\sigma}t_{imi'm'\sigma}\hat{O}^\dag_{i\alpha}\hat{O}_{i'\alpha'}\hat{f}^\dag_{im\sigma}\hat{f}_{i'm'\sigma}\,.
\end{multline*}
For the onsite $\epsilon_{im\sigma}$ terms, we have replaced $\hat f^\dag_{im\sigma} \hat f_{im\sigma} \hat O_{i\alpha}^\dag\hat O_{i\alpha}$ by the simpler $\hat f^\dag_{im\sigma} \hat f_{im\sigma}$ because even though $\hat O_{i\alpha}^\dag \hat O_{i\alpha}$ is not necessarily identity (unless $C_{i\alpha}=1)$, the two set of operators act identically on all the physical states of interest due to the fact that $\hat f_{im\sigma}$ annihilates the state with zero particles.  

The point of introducing the slave degree of freedom is that they track the number of electrons in the various orbitals and spin states on each site {\it and} the interaction Hamiltonian of Eq.~(\ref{eq:SlaterKanamori}) is essentially determined by these numbers.  Hence, one can write the most important parts of the interaction Hamiltonian solely in terms of the bosonic slave operators.

\subsection{Decoupling spinons and slaves}

Up to this point, our considerations have been for an exact solution of the interacting Hamiltonian which is impossible in practice.  To make progress, in slave-particle approaches one splits the problem into two separate and simpler pieces that are connected to each other via self-consistent averages of the relevant operators.  Specifically, the ground-state wave function of the system is approximated by a simple, separable product state $\ket{\Psi_f}\ket{\Phi_s}$ where $\ket{\Psi_F}$ is a spinon wave function and $\ket{\Phi_s}$ is a slave boson wave function.  Since the spinons and slaves are now decoupled and their number fluctuations are no longer locked in step, the constraint of Eq.~(\ref{eq:fmatchNconstraint}) can only be satisfied on average,
\begin{dmath}
\langle \sum_{m\sigma \in \alpha}\hat f^\dag_{im\sigma} \hat f_{im\sigma}\rangle_f=\langle \hat{N}_{i\alpha}\rangle_s\,.
\label{eq:numconstraint}
\end{dmath}
where the $f$ and $s$ subscripts denote averaging over the spinon $|\Psi_f\rangle$ and slave boson $|\Phi_s\rangle$ ground state wave functions, respectively.

{\it A priori}, one can make only few statements about when such a decoupling scheme is expected to be a good approximation. First, at zero interaction strength ($U=J=0$), the decoupled approach can reproduce the non-interacting band structure since the spinons alone can do this.  Second, if number fluctuations about the averages are small so that imposing Eq.~(\ref{eq:numconstraint}) is close to imposing Eq.~(\ref{eq:fmatchNconstraint}) , then we expect the decoupling to work well in describing ground-state averages; examples of phases with small or zero number fluctuations are narrow or Mott-like insulating bands.  Third, in infinite dimensions as well as in one dimension, due to the similarity of the main equations (see below) with the Gutzwiller approximation \cite{Florens2004,DeMedici2005,Bunemann1997,Bunemann2011}, one can say the two approaches should succeed or fail together. Unfortunately, in two or three dimensions --- which are cases of common interest --- there is no {\it a priori} way to know whether the approximation will be good or poor.

We note that when the decoupling approximation is performed and the constraint imposed only on average in Eq.~(\ref{eq:numconstraint}), we may decide to change $N_{min}$ and $N_{max}$ to allow additional unphysical states with negative or positive occupations.  For example, letting $N_{min}\rightarrow-\infty$ and $N_{max}\rightarrow+\infty$, which in turn makes $C_{i\alpha}$ irrelevant, yields the mean-field O(2) slave-rotor formalism used in previous work\cite{Florens2002, Florens2004}.  As  noted previously\cite{DeMedici2005}, allowing these unphysical states is not a major error in the limit of strong interactions since number fluctuations to unphysical occupancies are at any rate unlikely due to their large Coulombic penalties; however, for small interaction strengths, the unphysical state have significant weight in the wave function which creates incorrect behavior, e.g., improper behavior of the quasiparticle weight versus interaction strength.  At the other extreme, a separate slave boson for each spin+orbital combination $im\sigma$ gives $N_{min}=0$ and $N_{max}=1$ which is just the ``slave-spin'' formalism.\cite{DeMedici2005,Hassan2010}

With this separability assumption, the time-independent Schr\"odinger equation for the original system separates into two separate equations where the constraints of Eq.~(\ref{eq:numconstraint}) are enforced by Lagrange multipliers $h_{i\alpha}$ appearing in the two Hamiltonians.  In the remainder of this section, we discuss the simplest $U=U'$ and $J=0$ case for simplicity.  Full expressions involving $U$, $U'$ and $J$ for various slave boson choices follow after this section. The spinon Hamiltonian is
\begin{multline}
\hat{H}_f=\sum_{im\sigma}\epsilon_{im\sigma}\hat{f}_{im\sigma}^\dag\hat{f}_{im\sigma}\\
-\sum_{i\alpha}h_{i\alpha}\sum_{m\sigma\in\alpha}\hat{f}_{im\sigma}^\dag\hat{f}_{im\sigma}\\
-\sum_{ii'\alpha\alpha'} \langle\hat{O}^\dag_{i\alpha}\hat{O}_{i'\alpha'}\rangle_s\!
\sum_{\substack{m\sigma\in\alpha\\m'\sigma\in\alpha'}}
t_{imi'm'\sigma}  \hat{f}^\dag_{im\sigma}\hat{f}_{i'm'\sigma}\,.
\end{multline}
The spinons are coupled to the slave bosons via the average $\langle\hat{O}^\dag_{i\alpha}\hat{O}_{i'\alpha'}\rangle_s$ which renormalizes  spinon hoppings between sites $i$ and $i'$.  The spinon problem is one of non-interacting fermionic particles with spin.

The slave boson Hamiltonian takes the form
\begin{multline}
\hat{H}_{s}=\frac{U}{2}\sum_i\left[\left(\sum_{\alpha}\hat{N}_{i\alpha}\right)^2-\sum_{\alpha}\hat{N}_{i\alpha}\right]+\sum_{\alpha}h_{i\alpha}\hat{N}_{i\alpha}\\
-\sum_{ii'\alpha\alpha'}\left[
\sum_{\substack{m\sigma\in\alpha\\m'\sigma\in\alpha'}}t_{imi'm'\sigma}\langle \hat{f}^\dag_{im\sigma}\hat{f}_{i'm'\sigma} \rangle_f \right]
 \hat{O}^\dag_{i\alpha}\hat{O}_{i'\alpha'}
\end{multline}
where  the spinon average $\langle \hat{f}^\dag_{im\sigma}\hat{f}_{i'm'\sigma} \rangle_f$ renormalizes the slave boson hoppings.  The slave boson problem is one of interacting charged bosons without spin.

The original problem has been reduced to a set of paired problems that must be solved self-consistently.  The spinon and slave boson problems only communicate (i.e., are coupled) via averages which renormalize each other's hoppings.  
At this point, one must make some approximations in order to solve the interacting bosonic problem.  Typical approaches to date include single-site mean field approximations \citep{Florens2002, Florens2004}, multiple-site mean field \citep{Zhao2007}, approximation by sigma models to yield Gaussian integrals \citep{Florens2002, Florens2004} as well as a combination of using tight-binding parameters obtained using Wannier functions from DFT followed by a mean-field approximation \citep{Lau2013a}.

Separately, a procedure is needed to obtain the $C_{i\alpha}$. To this end, at $U=U'=J=0$, one chooses the $C_{i\alpha}$ to ensure that the spinon bands reproduce the original non-interacting band structure and associated  occupancies (i.e., fillings).  This means that the slave-boson expectations $\langle O_{i\alpha}^\dag O_{i'\alpha'}\rangle_s$ should be unity in order not to modify the spinon hoppings away from the original non-interacting hoppings  $t_{imi'm'\sigma}$.  The numbers $C_{i\alpha}$ and $h_{i\alpha}$ are determined by making $\langle O_{i\alpha}^\dag O_{i'\alpha'}\rangle_s$ unity as well as reproducing the non-interacting occupancies or fillings.
This actually requires us to solve the coupled slave and spinon problems at $U=U'=J=0$ self-consistently to obtain $C_{i\alpha}$ and $h_{i\alpha}$.  The values of $C_{i\alpha}$ are then held fixed from that point forth when turning on $U,U',J$ to non-zero values to self-consistently solve the actual interacting problem.

Prior to solving some model problems within our new framework, we provide more complete descriptions of a number of potential choices for the slave-boson model (i.e., the choice of $\alpha$) with full $U$, $U'$ and $J$ dependence.  Differing choices split the interaction terms $\hat H^i_{int}$ of Eq.~(\ref{eq:SlaterKanamori}) in different ways between the spinon and slave sectors.  This  opens the door to systematic comparison between the different types of treatments of correlations with the slave bosons.

\subsubsection{Number slave}

The simplest approach is to simply create a single slave boson on each site $i$ whose number operator $\hat N_i$ counts all the electrons on that site.  In other words, the label $\alpha$ contains all the $m\sigma$ orbitals on that site: it is superfluous so we can  write $\hat O_{i\alpha}=\hat O_i$.  Description of the physically allowed states will set $N_{min}=0$ while $N_{max}$ will be the maximum number of electrons allowed on that site: e.g., 10 for $d$ shells or 14 for $f$ shells.

In this case, the slave boson can only represent the $U$ term of the interaction in Eq.~(\ref{eq:SlaterKanamori}); all remaining interaction terms must be treated at the mean-field level in the spinon sector.   The slave Hamiltonian in this case is
\begin{multline}
\hat{H}_{s}=\frac{U}{2}\sum_i\left(\hat{N}_{i}^2-\hat{N}_{i}\right)+\sum_{i}h_{i}\hat{N}_{i}\\
-\sum_{ii'}\left[\sum_{mm'\sigma}t_{imi'm'\sigma}\langle \hat{f}^\dag_{im\sigma}\hat{f}_{i'm'\sigma}\rangle_f
\right]\hat{O}^\dag_{i}\hat{O}_{i'}
\end{multline}
while the spinon Hamiltonian contains all the remaining interaction terms at mean-field level:
\begin{multline}
\hat{H}_f= \frac{U'-U}{2}\sum_i\sum_{m\ne m'} \Big( n_{im} \hat n_{im'} + n_{im'} \hat n_{im}\\
 - \sum_{\sigma\sigma'}\left\{\rho_{im'\sigma'im\sigma}\hat f_{im'\sigma'}^\dag \hat f_{im\sigma} + \rho_{im\sigma im'\sigma'}\hat f_{im\sigma}^\dag \hat f_{im'\sigma'}\right\}\Big)\\
-\frac{J}{2}\sum_{i\sigma}\sum_{m\ne m'}\Big(n_{im\sigma}\hat n_{im'\sigma}+n_{im'\sigma}\hat n_{im\sigma}\\
- \rho_{im'\sigma'im\sigma}\hat f_{im'\sigma'}^\dag \hat f_{im\sigma}
- \rho_{im\sigma im'\sigma'}\hat f_{im\sigma}^\dag \hat f_{im'\sigma'}
\Big)\\
- \frac{J}{2}\sum_{i\sigma}\sum_{m\ne m'}\Big( \rho_{im\bar\sigma im\sigma}\hat{f}^\dag_{im'\bar\sigma}\hat{f}_{im'\sigma}
+ \rho_{im'\sigma im'\bar\sigma}\hat{f}^\dag_{im\sigma}\hat{f}_{im\bar\sigma}\\
- \rho_{im'\sigma im\sigma}\hat{f}^\dag_{im'\bar\sigma}\hat{f}_{im\bar\sigma}
- \rho_{im\bar\sigma im'\bar\sigma}\hat{f}^\dag_{im\sigma}\hat{f}_{im'\sigma}\\
+ \rho_{im'\bar\sigma im\sigma}\hat{f}^\dag_{im\bar\sigma}\hat{f}_{im'\sigma}
+ \rho_{im'\sigma im\bar\sigma}\hat{f}^\dag_{im\sigma}\hat{f}_{im'\bar\sigma}\\
- \rho_{im'\sigma im\sigma}\hat{f}^\dag_{im\bar\sigma}\hat{f}_{im'\bar\sigma}
- \rho_{im'\bar\sigma im\bar\sigma}\hat{f}^\dag_{im\sigma}\hat{f}_{im'\sigma}
\Big)
\\
+\sum_{im\sigma}\epsilon_{im\sigma}\hat{f}_{im\sigma}^\dag\hat{f}_{im\sigma}-\sum_{i}h_{i}\hat n_i\\
-\sum_{ii'} \langle\hat{O}^\dag_{i}\hat{O}_{i'}\rangle_s\!
\sum_{mm'\sigma}t_{imi'm'\sigma}  \hat{f}^\dag_{im\sigma}\hat{f}_{i'm'\sigma}\,.
\end{multline}
In the derivation of the expression for the above spinon Hamiltonian $\hat H_f$, we have used the definition of the one-particle density matrix
\[
\rho_{ba} = \langle \hat f^\dag_a \hat f_b\rangle_f\,,
\]
the standard mean-field contraction of four particle operators into two-particle operators weighed by averages 
\[
 \hat f^\dag_a \hat f^\dag _b \hat f_c \hat f_d \approx 
 \rho_{da} \hat f^\dag_b \hat f_c  - \rho_{ca}\hat f^\dag _b  \hat f_d
+ \rho_{cb}\hat f^\dag_a\hat f_d  - \rho_{db}\hat f^\dag _a  \hat f_c\,,
\]
and the average occupations
\[
n_{im\sigma} = \rho_{im\sigma im\sigma} \ \ , \ \ n_{im} = \sum_\sigma n_{im\sigma}\,.
\]
This approach has the simplest slave Hamiltonian and the most complex spinon Hamiltonian.  This is because the number-only slave boson can only describe the simplest $U$ part of the interaction; the remaining terms involving $U'$ and $J$ must be handled at mean-field level by the spinons.  As mentioned above, the physical range for the occupation numbers of the number slave $\hat N_i$ is from zero to the physically allowed $N_{max}$ for that site.  However, we can  decrease $N_{min}$ below zero and $N_{max}$ above the physical value if desired: in the limit where the range of occupancies allowed is very large, we automatically recover the O(2) slave-rotor method. 

\subsubsection{Orbital slave}

A more fine-grained model is to count the number of electrons in each spatial orbital $m$ separately with a slave boson.  We call this the orbital slave method.  Here the index $\alpha$ labels a specific spatial orbital $m$ and ranges over the two spin directions for that orbital:  we have $\hat O_{im}$ for the raising/lowering operator and $\hat N_{im}$ for the particle count slave operators.  The slave sector can now directly describe more of the interaction terms:
\begin{multline}
\hat{H}_{s}=\frac{U}{2}\sum_i\left(\left(\sum_{m}\hat{N}_{im}\right)^2-\sum_{m}\hat{N}_{im}\right)\\
+\frac{U'-U}{2}\sum_i\sum_{m\ne m'}\hat{N}_{im}\hat{N}_{im'} + \sum_{i}\sum_{m}h_{im}\hat{N}_{im}
\\
-\sum_{ii'mm'}\left[ \sum_\sigma t_{imi'm'\sigma}\langle \hat{f}^\dag_{im\sigma}\hat{f}_{i'm'\sigma} \rangle_f \right]\hat{O}^\dag_{im}\hat{O}_{i'm'}
\end{multline}
and the spinon Hamiltonian is less complex than the previous case as it only has the $J$ terms at mean-field level:
\begin{multline}
\hat{H}_f=
-\frac{J}{2}\sum_{i\sigma}\sum_{m\ne m'}\Big(n_{im\sigma}\hat n_{im'\sigma}+n_{im'\sigma}\hat n_{im\sigma}\\
- \rho_{im'\sigma'im\sigma}\hat f_{im'\sigma'}^\dag \hat f_{im\sigma}
- \rho_{im\sigma im'\sigma'}\hat f_{im\sigma}^\dag \hat f_{im'\sigma'}\Big)\\
- \frac{J}{2}\sum_{i\sigma}\sum_{m\ne m'}\Big( \rho_{im\bar\sigma im\sigma}\hat{f}^\dag_{im'\bar\sigma}\hat{f}_{im'\sigma}
+ \rho_{im'\sigma im'\bar\sigma}\hat{f}^\dag_{im\sigma}\hat{f}_{im\bar\sigma}\\
- \rho_{im'\sigma im\sigma}\hat{f}^\dag_{im'\bar\sigma}\hat{f}_{im\bar\sigma}
- \rho_{im\bar\sigma im'\bar\sigma}\hat{f}^\dag_{im\sigma}\hat{f}_{im'\sigma}\\
+ \rho_{im'\bar\sigma im\sigma}\hat{f}^\dag_{im\bar\sigma}\hat{f}_{im'\sigma}
+ \rho_{im'\sigma im\bar\sigma}\hat{f}^\dag_{im\sigma}\hat{f}_{im'\bar\sigma}\\
- \rho_{im'\sigma im\sigma}\hat{f}^\dag_{im\bar\sigma}\hat{f}_{im'\bar\sigma}
- \rho_{im'\bar\sigma im\bar\sigma}\hat{f}^\dag_{im\sigma}\hat{f}_{im'\sigma}
\Big)
\\
+\sum_{im\sigma}\epsilon_{im\sigma}\hat{f}_{im\sigma}^\dag\hat{f}_{im\sigma}-\sum_{i}\sum_m h_{im}\hat n_{im}\\
-\sum_{ii'mm'} \langle\hat{O}^\dag_{im}\hat{O}_{i'm'}\rangle_s\!
\sum_{\sigma}t_{imi'm'\sigma}  \hat{f}^\dag_{im\sigma}\hat{f}_{i'm'\sigma}\,.
\end{multline}

\subsubsection{Spin slave}

An alternative fine-graining  is to have two slave bosons per site that count spin up and spin down electrons separately but with no orbital differentiation.  Namely, $\alpha$ labels a spin state $\sigma$ but ranges over all spatial orbitals.  Hence, we have $\hat O_{i\sigma}$ and $\hat N_{i\sigma}$ for our slave operators.  The slave-boson Hamiltonian is
\begin{multline}
\hat{H}_{s}=\frac{U}{2}\sum_i\left(\left(\sum_{\sigma}\hat{N}_{i\sigma}\right)^2-\sum_{\sigma}\hat{N}_{i\sigma}\right)\\
-\frac{J}{2}\sum_{\sigma}\left(\hat N_{i\sigma}^2-\hat N_{i\sigma}\right) \\
 + \sum_{i}\sum_{\sigma}h_{i\sigma}\hat{N}_{i\sigma}
\\
-\sum_{ii'\sigma}\left[ \sum_{mm'} t_{imi'm'\sigma}\langle \hat{f}^\dag_{im\sigma}\hat{f}_{i'm'\sigma} \rangle_f \right]\hat{O}^\dag_{i\sigma}\hat{O}_{i'\sigma}
\end{multline}
while the spinon Hamiltonian is
\begin{multline}
\hat{H}_f= \frac{U'-U}{2}\sum_i\sum_{m\ne m'} \Big( n_{im} \hat n_{im'} + n_{im'} \hat n_{im}\\
 - \sum_{\sigma\sigma'}\left\{\rho_{im'\sigma'im\sigma}\hat f_{im'\sigma'}^\dag \hat f_{im\sigma} + \rho_{im\sigma im'\sigma'}\hat f_{im\sigma}^\dag \hat f_{im'\sigma'}\right\}\Big)\\
- \frac{J}{2}\sum_{i\sigma}\sum_{m\ne m'}\Big( \rho_{im\bar\sigma im\sigma}\hat{f}^\dag_{im'\bar\sigma}\hat{f}_{im'\sigma}
+ \rho_{im'\sigma im'\bar\sigma}\hat{f}^\dag_{im\sigma}\hat{f}_{im\bar\sigma}\\
- \rho_{im'\sigma im\sigma}\hat{f}^\dag_{im'\bar\sigma}\hat{f}_{im\bar\sigma}
- \rho_{im\bar\sigma im'\bar\sigma}\hat{f}^\dag_{im\sigma}\hat{f}_{im'\sigma}\\
+ \rho_{im'\bar\sigma im\sigma}\hat{f}^\dag_{im\bar\sigma}\hat{f}_{im'\sigma}
+ \rho_{im'\sigma im\bar\sigma}\hat{f}^\dag_{im\sigma}\hat{f}_{im'\bar\sigma}\\
- \rho_{im'\sigma im\sigma}\hat{f}^\dag_{im\bar\sigma}\hat{f}_{im'\bar\sigma}
- \rho_{im'\bar\sigma im\bar\sigma}\hat{f}^\dag_{im\sigma}\hat{f}_{im'\sigma}
\Big)
\\
+\sum_{im\sigma}\epsilon_{im\sigma}\hat{f}_{im\sigma}^\dag\hat{f}_{im\sigma}-\sum_{i}\sum_m h_{im}\hat n_{im}\\
-\sum_{ii'mm'} \langle\hat{O}^\dag_{im}\hat{O}_{i'm'}\rangle_s\!
\sum_{\sigma}t_{imi'm'\sigma}  \hat{f}^\dag_{im\sigma}\hat{f}_{i'm'\sigma}\,.
\end{multline}

\subsubsection{Spin+orbital slave}

This approach represents maximum fine-graining whereby we use a slave boson for each spin+orbital combination.  Thus the index $\alpha$ now represents a full set of quantum numbers $m\sigma$ so we have $\hat O_{im\sigma}$ and $\hat N_{im\sigma}$ for the slave operators.  The physically allowed occupancies are 0 and 1 which is isomorphic to a pseudo-spin.  For this reason, the name used for this approach in the literature is the ``slave-spin'' method \citep{DeMedici2005,Hassan2010}.  However, given the possible confusion this term creates between the real electron spin as well as the difficulty of using such a name unambiguously in our generalized formalism, we  prefer the more explicit name  ``spin+orbital slave'' where the spin refers to the physical electron spin.

In this approach, we can describe the maximum number of interaction terms in the slave Hamiltonian: 
\begin{multline}
\hat{H}_s =\frac{U}{2}\sum_i \left( \left(\sum_{m\sigma}\hat{N}_{im\sigma}\right)^2-\sum_{m\sigma}\hat{N}_{im\sigma}\right)\\
+\frac{U'-U}{2}\sum_{m\neq m'}\left(\sum_\sigma \hat N_{im\sigma}\right)\left(\sum_{\sigma'}\hat{N}_{im'\sigma'}\right)\\
-\frac{J}{2}\sum_{\sigma}\sum_{m\neq m'}\hat{N}_{im\sigma}\hat{N}_{im'\sigma}\\
+ \sum_{i}\sum_{m\sigma}h_{im\sigma}\hat{N}_{im\sigma}
\\
-\sum_{ii'mm'\sigma} t_{imi'm'\sigma}\langle \hat{f}^\dag_{im\sigma}\hat{f}_{i'm'\sigma} \rangle_f \hat{O}^\dag_{im\sigma}\hat{O}_{i'm'\sigma}\,.
\end{multline}
The corresponding spinon Hamiltonian still contains the spin flip and pair-hopping terms:
\begin{multline}
\hat{H}_f= 
- \frac{J}{2}\sum_{i\sigma}\sum_{m\ne m'}\\
\Big( \rho_{im\bar\sigma im\sigma}\hat{f}^\dag_{im'\bar\sigma}\hat{f}_{im'\sigma}
+ \rho_{im'\sigma im'\bar\sigma}\hat{f}^\dag_{im\sigma}\hat{f}_{im\bar\sigma}\\
- \rho_{im'\sigma im\sigma}\hat{f}^\dag_{im'\bar\sigma}\hat{f}_{im\bar\sigma}
- \rho_{im\bar\sigma im'\bar\sigma}\hat{f}^\dag_{im\sigma}\hat{f}_{im'\sigma}\\
+ \rho_{im'\bar\sigma im\sigma}\hat{f}^\dag_{im\bar\sigma}\hat{f}_{im'\sigma}
+ \rho_{im'\sigma im\bar\sigma}\hat{f}^\dag_{im\sigma}\hat{f}_{im'\bar\sigma}\\
- \rho_{im'\sigma im\sigma}\hat{f}^\dag_{im\bar\sigma}\hat{f}_{im'\bar\sigma}
- \rho_{im'\bar\sigma im\bar\sigma}\hat{f}^\dag_{im\sigma}\hat{f}_{im'\sigma}
\Big)
\\
+\sum_{im\sigma}\epsilon_{im\sigma}\hat{f}_{im\sigma}^\dag\hat{f}_{im\sigma}-\sum_{i}\sum_m h_{im}\hat n_{im}\\
-\sum_{ii'mm'} \langle\hat{O}^\dag_{im}\hat{O}_{i'm'}\rangle_s\!
\sum_{\sigma}t_{imi'm'\sigma}  \hat{f}^\dag_{im\sigma}\hat{f}_{i'm'\sigma}\,.
\end{multline}

We mention that in prior work \citep{Hassan2010}, the spin flip and pair hopping terms  were argued to be well treated in the slave-particle sector instead.  Namely, they were removed from the spinon Hamiltonian and the following terms were added to the spin+orbital slave Hamiltonian:
\begin{multline}
-J\sum_{m \ne m'} \big( \hat S^{+}_{im\uparrow}\hat S^{-}_{im\downarrow}\hat S^{+}_{im'\downarrow}\hat S^{-}_{im'\uparrow}\\
+\hat S^{+}_{im\uparrow}\hat S^{+}_{im\downarrow}\hat S^{-}_{im'\uparrow}\hat S^{-}_{im'\downarrow}+h.c. \big)
\label{eq:adhocsfph}
\end{multline}
where the $\hat S$ operators in the number basis are
\begin{equation}
\hat S^{+}= \left( \begin{array}{cc}
0 & 0 \\
1 & 0 \end{array} \right)  \ \ \ \ , \ \ \ 
\hat S^{-}= \left( \begin{array}{cc}
0 & 1 \\
0 & 0 \end{array} \right)\,.
\end{equation}
While such an {\it ad hoc} approach is not the strictly theoretically consistent way to split operators between the spinon and slave boson sectors, in practice it does reproduce the desired behavior of the spin flip and pair hopping terms in the slave boson sector and does not introduce any numerical difficulties.

Our approach provides some insights into this inconsistency issue while simultaneously easing some technical problems that can arise in the spin+orbital slave approach.  Part of the inconsistency is that the {\it ad hoc} $\hat S^\pm$ slave operators are not the same as the $\hat{O}$ operators (the only way to make them the same is the extreme choice $C_{im\sigma}=0$).  For example, at half-filling when $C_{im\sigma}=1$,  $\hat{O}$ and $\hat{O}^\dag$ are the same operator and equal the $\hat S_x$ Pauli matrix, so one can not use $\hat{O}$ and $\hat{O}^\dag$ to represent any sensible representation of the the spin-flip or pair-hopping interaction terms.  If we choose to include some unphysical states, however, things become different.  For example, if we widen the range of the spin+orbital slave boson occupancies from $\{0,1\}$ to $\{-1,0,1,2\}$ then $\hat{O}$ and $\hat{O}$ become different.  One can then write a more natural interaction term of the form
\begin{multline}
-J\sum_{m \ne m'} \big( \hat O^{\dag}_{im\uparrow}\hat O_{im\downarrow}\hat O^{\dag}_{im'\downarrow}\hat O_{im'\uparrow}\\
+\hat O^{\dag}_{im\uparrow}\hat O^{\dag}_{im\downarrow}\hat O_{im'\uparrow}\hat O_{im'\downarrow}+h.c. \big)
\end{multline}
that only uses the slave boson $\hat{O}$ operators.

Separately, enlarging the set of occupancies beyond $\{0,1\}$ has some technical advantages.  When the occupancies are limited to $\{0,1\}$, then we have the analytical form\cite{Hasan2010} given by $C_{im\sigma}=[n_{im\sigma}(1-n_{im\sigma})]^{-1/2}$.  For occupancies nearing the extremes ($n_{im\sigma}\rightarrow0$ or $1$), the $C_{im\sigma}$ become very large, the $O_{im\sigma}$ matrices become ill-behaved, and the numerical algorithm using them becomes difficult to stabilize.  Permitting a wider range of occupancies such as $\{-1,0,1,2\}$ makes the $C_{im\sigma}$ have reasonable values and the numerical procedure is well behaved.  In the large interaction limit, the addition of the unphysical states (e.g., occupancies $-1$ and $2$ in this case) is not a major error since number fluctuations are suppressed; the main problem is in the weak interaction regime where this enlargement of the occupancy basis provides numerical stability but can produce quantitative errors.

\section{Mean-Field Tests}
\label{sec:tests}
We now proceed to describe computational results based on a simple single-site, paramagnetic, nearest-neighbor, mean-field solution of the slave Hamiltonian at half filling.  This will permit us to both reproduce prior literature as well as to compare various slave Hamiltonians to each other.  

To do so, we shift the local interaction energies so that they are zero at half filling, i.e. when $n_{im\sigma}=1/2$.  We also make the standard choice $U'=U-2J$.  The local interaction term (ignoring for the moment the spin flip and pair-hopping terms) takes the form from prior work\cite{DeMedici2005}:
\begin{dmath}
\hat{H}_{int}^i=\frac{U-2J}{2}(\hat{n}_{i}-n^i_{orb})^2+J\sum_{m}(\hat{n}_{im}-1)^2-\frac{J}{2}\sum_{\sigma}(\hat{n}_{i\sigma}-n^i_{orb}/2)^2
\label{eq:inthalffill}
\end{dmath}
where $n^i_{orb}$ is the number of localized correlated spatial orbitals on site $i$.

In the single-site mean-field approximation, we will be solving for a single site self-consistently coupled to an averaged bath of bosons on the nearest neighbor sites.  Our assumptions ensure that all sites are identical with no spin polarization. Furthermore, to connect to the literature, we  further assume that in the multi-orbital case there are only non-zero hoppings between nearest neighbor orbitals with the same $m$ index.  With all these assumptions, it is easy to see that $C_{i\alpha}=1$ is the choice that gives half-filling for the slave problem at $U=U'=J=0$.  In addition, we can set the Lagrange multipliers $h_{i\alpha}=0$ since we have set the half-filling energy to be zero.  The density matrix elements $\langle \hat f_{im\sigma}^{\dagger}\hat f_{i'm'\sigma} \rangle_f$ that renormalize the slave boson hoppings will be spin and site independent and will be non-zero only when $m=m$'.  Hence, they can be absorbed into the definition of the hopping elements $t_{imi'm'\sigma}$.  The density of states for a spinon band is taken to be the standard semicircular one
\[
D(E) = \left\{\begin{matrix}{c}\frac{\sqrt{4t^2-E^2}}{2\pi t^2} \ \ \ \mbox{if } |E|<2|t|\\0 \ \ \ \mbox{else}\end{matrix}\right. \,.
\]

We begin with  $J=0$.  The slave Hamiltonian  is 
\begin{multline}
H_{s}= \frac{U}{2}\sum_i \left( \sum_{\alpha}\hat{N}_{i\alpha}-n_{orb}\right)^2 \\
- \sum_{i\alpha}\sum_{m\in\alpha}\left(\hat O_{i\alpha}t^{eff}_m + \hat O^\dag_{i\alpha}t^{eff}_m\right)
\end{multline}
where the effective hoping for spatial orbital $m$ is
\begin{equation}
t^{eff}_m = \sum_{i'\alpha'}\sum_{m'\sigma\in\alpha'}
t_{imi'm'\sigma}\langle \hat O_{\alpha'} \rangle_{s}\,.
\end{equation}
The simple form of this Hamiltonian makes it easy to directly read off the quasiparticle weight renormalization $Z_\alpha$ which narrows the spinon bands:
\begin{equation}
Z_{\alpha}=\langle \hat{O}_{\alpha} \rangle_s ^2 \,.
\end{equation}
When $Z_\alpha=0$, a Mott insulator is realized in such a simple single-site model \citep{Brinkman1970}.
We solve the problem self-consistently for different slave models. Since at half-filling the Lagrange multipliers $h_{i\alpha}=0$, all that is required to solve the spinon problem is to renormalize each spinon band width (i.e., hopping) by the appropriate $Z_\alpha$ factor.

\subsection{Single-band  Mott transition}

\begin{figure}
\centering
\includegraphics[width=0.5\textwidth]{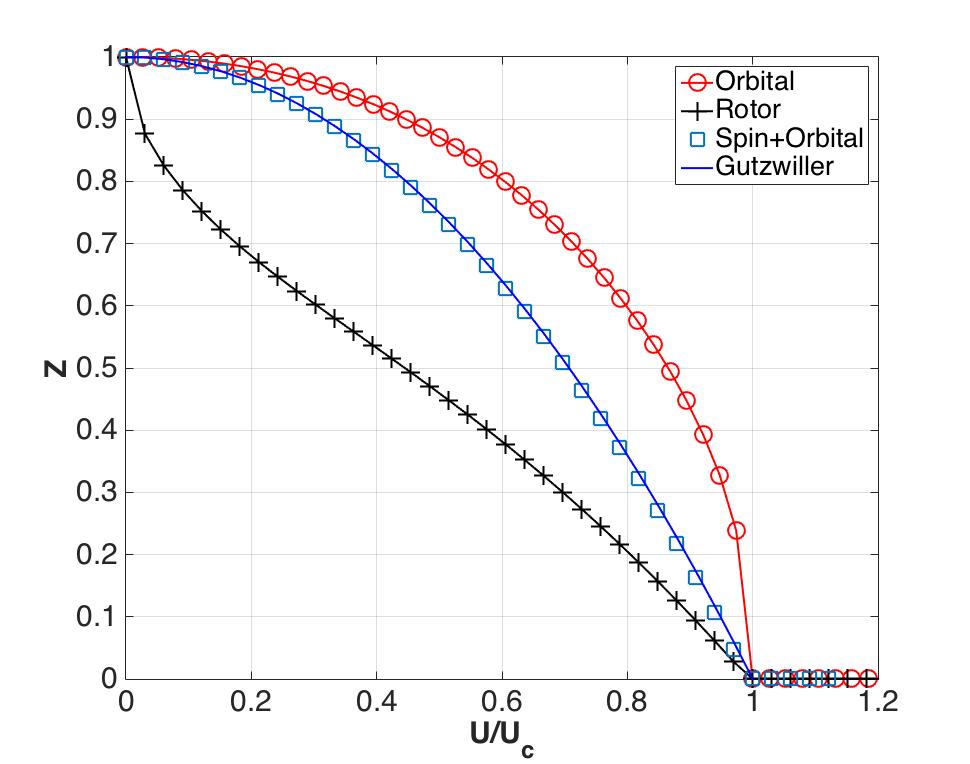}
\caption{Quasiparticle weight $Z$ as a function of $U/U_c$ for different slave-particle models for the paramagnetic single-band Hubbard at half filling. $U_c$ is the critical value of $U$ when $Z=0$, i.e., the Mott transition, for each model.  The black squares show slave rotor results, the blue line is the Gutzwiller approximation results which for this model are  the same as the spin+orbital slave (``slave-spin'') results in blue crosses, and the red circles show the orbital slave results (identical to the number slave). We note that the orbital-slave Hilbert space is very small, so that it does not agree with the rotor, unlike the two-band slave number. }
\label{fig:singleband}
\end{figure}

We begin with a single-band model where there is one spatial orbital per site. Figure~\ref{fig:singleband} compares various slave models based on the dependence of $Z$ on $U$.  Specifically, we compare the slave rotor model (allowed occupancies from $-\infty$ to $+\infty$), the orbital slave model (allowed occupancies 0, 1, or 2) which here is identical to the number slave model, the spin+orbital slave (``slave-spin'') model (allowed occupancies 0 or 1) and the Gutzwiller approximation where $Z_{Gutzwiller}=1-(U/U_c)^2$.  

For this system, the Gutzwiller and spin+orbital slave methods predict exactly the same results, as noted previously.\citep{DeMedici2005}  In fact, the spin+orbital slave model, at half-filling for a single orbital per site at the single-site mean field level, can be shown to be isomorphic to the Gutzwiller approximation as well as to the Kotliar-Ruckenstein model as described by Bunemann.\citep{Bunemann2011} This shows that, beyond their utility as mathematical models, such slave-boson methods can parallel and help understand other approaches that originate from apparently different sets of many-body approximations. 

The slave-rotor method has an aberrant behavior for small $U$.  Specifically, $Z$ for the slave-rotor method has the small $U$ expansion
\begin{equation}
Z_{rotor} = 1 - O(\sqrt{U/t_{eff}})\,.
\end{equation}
The reason for this behavior is due to the unbounded number states permitted in the slave rotor model.  Specifically, in the number basis the slave-rotor problem corresponds to an infinite one dimensional lattice labeled by $N_i$, with hoppings $t^{eff}$ between neighboring sites, and with a quadratic potential $UN_i^2/2$.  For small $U$, the ground state of this problem will be spread over many sites so that we can take the continuum limit.  The problem turns into the textbook one dimensional harmonic oscillator with mass $1/(2t_{eff})$ and spring constant $U$.  The ground state wave function $\psi(N_i)$ is a Gaussian, and $\langle O\rangle_s=\sum_n \psi(n)\psi(n-1)$ can be computed.  Expansion in $U$ then gives the above form.

In reality, however, perturbation theory guarantees that quasiparticle weights are modified starting at second order in the interaction strength:
\begin{equation}Z=1-O((U/t^{eff})^2)\,.
\end{equation}
The slave-rotor fails since for small $U$ it spreads the wave function over a large number of unphysical states.  What this means is that one would incorrectly overestimate the importance of electronic correlations at weak interaction strengths when using the slave-rotor method.  In this view, our orbital and number slave methods may be viewed as corrected rotors which are restricted to the appropriate finite set of physical states.  Finally,  
Figure~\ref{fig:singleband} illustrates that slave methods employing finite slave Hilbert spaces all automatically correct the small $U$ behavior.  

\subsection{Isotropic two-band Mott transition}
\begin{figure}[h!]
\centering
\includegraphics[width=0.5\textwidth]{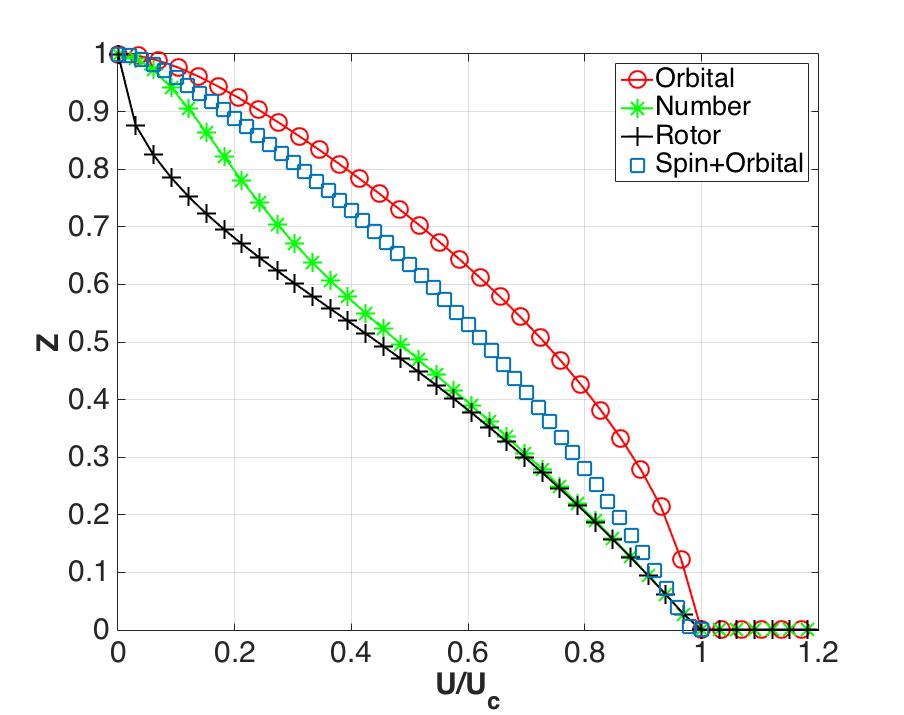}
\caption{Quasiparticle weight $Z$ as a function of $U/U_c$ for different slave-particle models for a degenerate paramagnetic two-band Hubbard model at half filling.    }
\label{fig:twodegband}
\end{figure}

Next, we consider a two-band degenerate Hubbard model.  Figure~\ref{fig:twodegband} displays the results.  We note that the two band $e_g$ model is of physical relevance as the slave-rotor has shown itself to be of use in $e_g$ nickelate systems within a $pd$ model\cite{Lau2013a}. For this particular degenerate case with high symmetry, the spin slave and orbital slave models turn out to be identical since each posits two slave particles each with the allowed occupations 0, 1, or 2. We note that, in this case, the slave rotor and number slave become very similar for large $U$:  once slave number fluctuations of $N_i$ are small, the size of the slave Hilbert space becomes irrelevant.

\subsection{Anisotropic Orbital-Selective Mott Transition}
We present mean-field calculations exemplifying the orbital-selective Mott transition in an anisotropic two band model with paramagnetic solution and at half filling.  We take spatial orbital $m=1$ to have the larger hopping $t_1$ while $m=2$ has the smaller hopping $t_2$.  Hence, $t_2/t_1$ specifies the degree of anisotropy.

The first slave model for this system is the spin+orbital method which has been used previously\cite{DeMedici2005,Hassan2010}: each slave boson has allowed occupancies 0 or 1. The second model is to forgo the explicit spin degree of freedom in the slave description and to employ the orbital slave model where each slave boson has allowed occupancies 0, 1, and 2.  The comparison tests the importance of explicit treatment of spin in the electronic correlations for such a system.  We will focus on the Orbital-Selective Mott Transition (OSMT) when one orbital has a finite bandwidth and is metallic while the other has undergone a Mott insulating transition and is localized.

\begin{figure}[h!]
\centering
\includegraphics[width=0.35\textwidth]{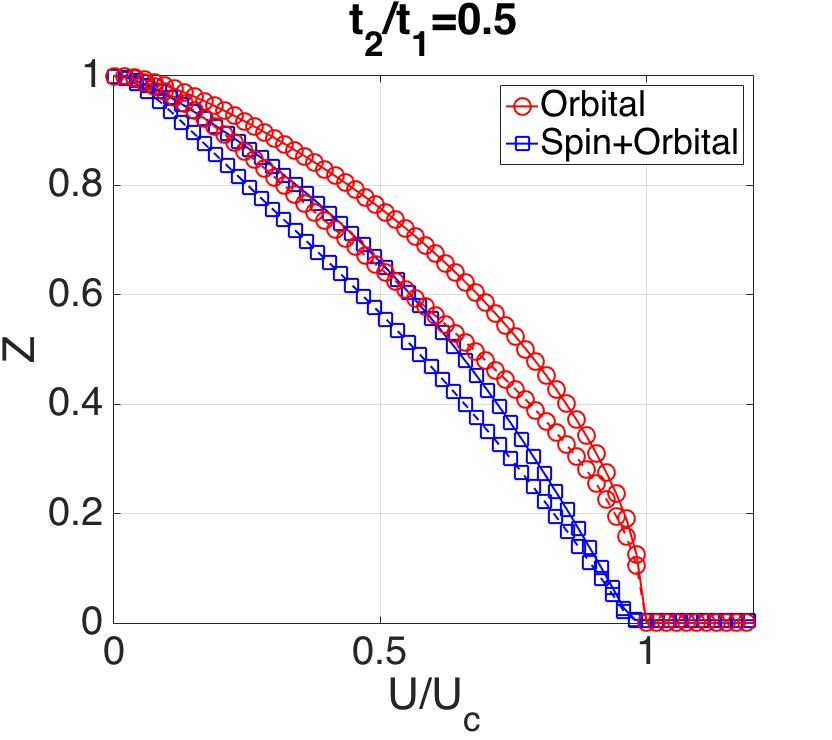}
\includegraphics[width=0.35\textwidth]{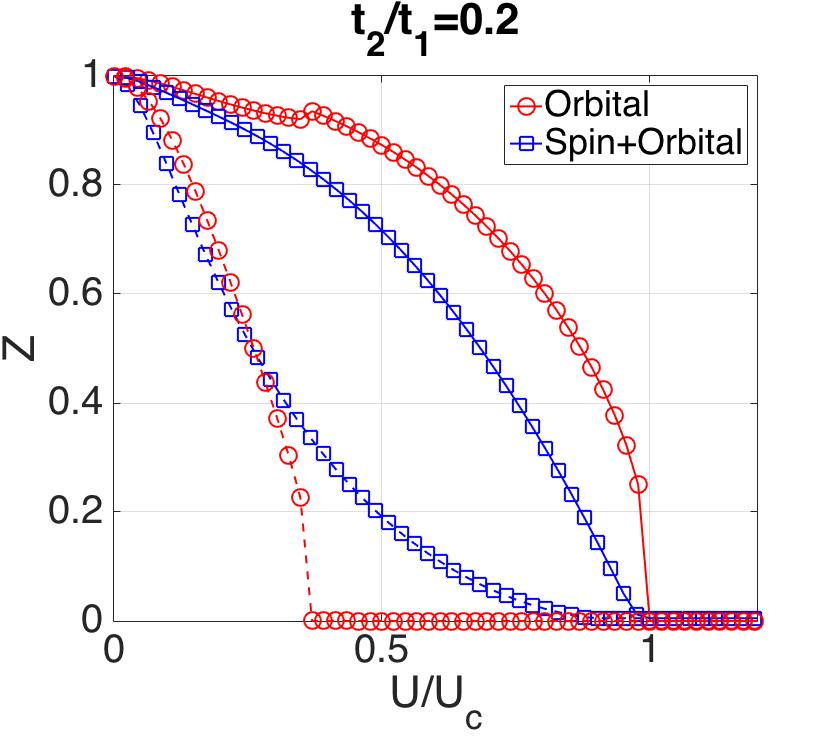}
\includegraphics[width=0.35\textwidth]{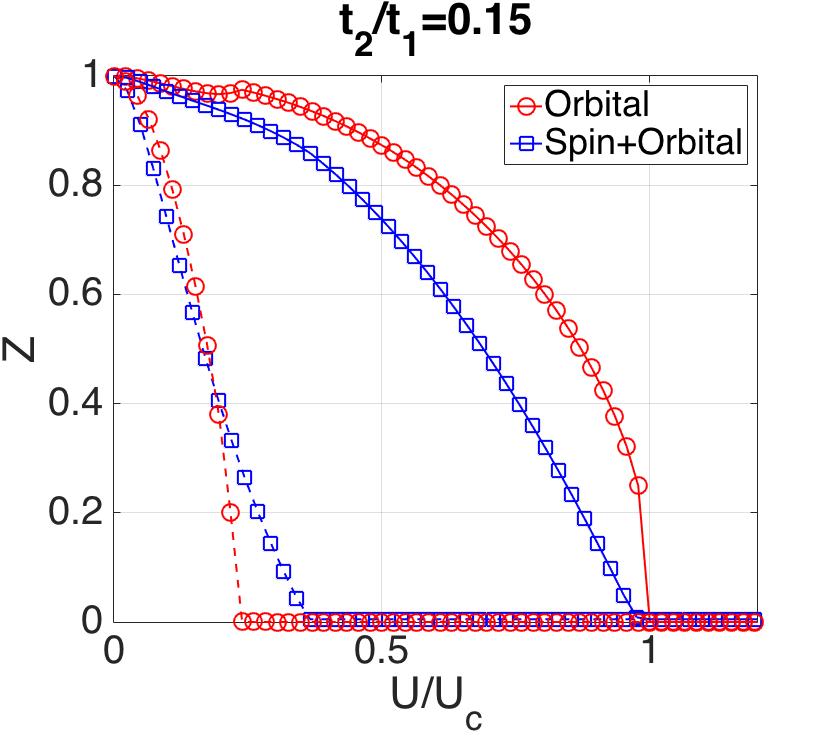}
\caption{Quasiparticle weights for the paramagnetic anisotropic two-band single-site Hubbard model at half filling as predicted by the orbital+spin slave model (blue squares) and the orbital slave model (red circles) at $J=0$ for three $t_2/t_1$ ratios.  In each plot, the $Z$ value for the first orbital with larger hopping $t_1$ is denoted by symbols and a continuous line while for the second orbital symbols and a dashed line are used.  An OSMT occurs when the two $Z$ do not go to zero at the same $U$ value: orbital slave (red circles) in the center plot and both slave models in the lower plot. }
\label{fig:z1z2vsU}
\end{figure}

We begin with $J=0$.  Figure~\ref{fig:z1z2vsU} illustrates the behavior of the renormalization factor $Z$ for both bands versus $U$ for three different $t_2/t_1$ ratios within the two slave particle models.  An OSMT occur for small enough $t_2/t_1$ ratio but the critical value depends on the type of slave model.  For the  orbital slave model, we find that OSMT occurs when $t_2/t_1<0.25$ while for spin+orbital slave we must have a slightly smaller value of $t_2/t_1<0.2$.

We now consider $J>0$. We continue to treat the spinon problem as that of a simple, paramagnetic, half-filled tight-binding model with two separate bands with each hopping renormalized by the appropriate $\langle \hat O_\alpha\rangle_s$. For the  orbital slave model, we can only include the first two terms of Eq.~(\ref{eq:inthalffill}) due to the lack of an explicit spin label in the slave description.  Thus we will compare the orbital slave and spin+orbital slave using the same interaction term
\begin{dmath}
\hat{H}_{int}^i=\frac{U-2J}{2}(\hat{N}_{i}-2)^2+J\sum_{m}(\hat{N}_{im}-1)^2\,.
\label{eq:hint2terms}
\end{dmath}
It is clear from the above two interaction terms that, for fixed $U$, $J>0$ permits larger orbital independent number fluctuations (i.e., it reduces the correlation effect of this mode) since $U'=U-2J$ becomes smaller in the first term.  However, the second $+J$ term simultaneously punishes intra-orbital number fluctuations and thus enhances intra-orbital correlation effects which in turn favors an OSMT.  

The phase diagram as a function of $t_2/t_1$ and $J$ for this system in shown in Figure~\ref{fig:osmtphase}.  The boundaries shown separate regions where OSMT occurs (above the boundaries) from where a standard Mott transition occurs (below the boundaries).  The figure confirms the fact that increasing $J$  favors OSMT.  Qualitatively, the orbital slave and spin+orbital slave show very similar behavior:  they have a critical $t_2/t_1$ at $J=0$ between $0.2-0.25$ for OSMT to occur, and then with increasing $J$ the critical $t_2/t_1$ becomes larger so less anisotropy is needed to drive an OSMT, as observed previously in DMFT \citep{Koga2004} and spin+orbital slave calculations\citep{DeMedici2005}. 

We have also considered the case where we permit the orbital slave model to have unlimited occupations:  namely, we have a two rotor model (one for each orbital occupation).  In this case, we find that no OSMT is possible when $J=0$ for any bandwidth ratio $t_2/t_1$.  This result is similar to previous DMFT\citep{Koga2004,DeMedici2005}, which found that a finite $J$ is needed in order to have an OSMT.  However, it disagrees with the results of previous orbital+spin slave results\citep{DeMedici2005} as well as our  results above where we find that a small enough bandwidth ratio $t_2/t_1$ makes an OSMT possible even for $J=0$.  These differences further illustrate the need for multiple models and cross verification when describing a possible OSMT  for real materials which have complex band structures (e.g., the three-band Ca$_{2-x}$Sr$_x$RuO$_4$  system\citep{Anisimov2002}).

Prior work\citep{DeMedici2005} has shown that the presence of the Hund's term
\begin{multline}
-\frac{J}{2} \sum_\sigma\sum_{m\ne m'} (\hat n_{m\sigma}-1/2)(\hat n_{m'\sigma}-1/2) \\=-\frac{J}{2}\sum_{\sigma}(\hat{n}_{i\sigma}-1)^2\ \ \ \ \,.
\end{multline}
 makes OSMT slightly more difficult to achieve as it increases inter-orbital $m\ne m'$ correlations by favoring spin pairing between different orbitals but does not aid intra-orbital correlations.  Separately,  adding the spin-flip and pair-hopping terms makes OSMT easier to achieve\citep{DeMedici2005}.

Although not directly relevant to our main focus, for completeness we include a final comparison based on a fixed slave model with various combination of interaction terms.  We choose the spin+orbital orbital model and then choose to include different interaction terms in the slave-particle Hamiltonian.  The first choice is the interaction terms used above in Eq.~({\ref{eq:hint2terms}).  The second choice is to add the Hund's term:
\begin{dmath}
\hat{H}_{int}^i=\frac{U-2J}{2}(\hat{N}_{i}-n^i_{orb})^2+J\sum_{m}(\hat{N}_{im}-1)^2-\frac{J}{2}\sum_{\sigma}(\hat{N}_{i\sigma}-1)^2\,.
\label{eq:Ham2}
\end{dmath}
Prior work\citep{DeMedici2005} has shown that the presence of the Hund's term
makes OSMT  more difficult to achieve as it increases inter-orbital correlations by favoring spin-pairing among different orbitals but does not enhance intra-orbital correlations.

The third choice is to add the spin-flip and pair-hopping terms as per the {\it ad hoc} method of Eq.~(\ref{eq:adhocsfph}):
\begin{dmath}
\hat{H}_{int}^i=\frac{U-2J}{2}(\hat{N}_{i}-2)^2+J\sum_{m}(\hat{N}_{im}-1)^2-\frac{J}{2}\sum_{\sigma}(\hat{N}_{i\sigma}-1)^2
\\-J\sum_{m \ne m'} \big( \hat S^{+}_{im\uparrow}\hat S^{-}_{im\downarrow}\hat S^{+}_{im'\downarrow}\hat S^{-}_{im'\uparrow}\\
+\hat S^{+}_{im\uparrow}\hat S^{+}_{im\downarrow}\hat S^{-}_{im'\uparrow}\hat S^{-}_{im'\downarrow}+h.c. \big)\,.
\label{eq:Ham3}
\end{dmath}
Adding these spin-flip and pair-hopping terms makes OSMT easier to achieve\citep{DeMedici2005}.

Phase diagrams for the second and third choices above are available in the literature\citep{DeMedici2005} and are reproduced in Figure~\ref{fig:osmtphasehint} which also includes the results of the first choice as well. We note that only including the intra-orbital terms (first choice) or all terms (third choice) leads to essentially the same phase diagram.  However, excluding the spin-flip and pair-hopping terms (second choice) makes it harder to achieve an OSMT phase: one can not achieve an OSMT for any reasonable $J$ once the bandwidth ratio $t_2/t_1$ exceeds $\approx0.6$. The physics behind this progression is as follows.  Starting with $J=0$ and a relatively large $U$, the ground-state basically contains only states which are half-filled and have a total of two electrons per site (there are six such states).  Adding the intra-orbital term (first choice) with $J>0$ then further restricts us to the four states with only one electron per orbital (but with no preference for spin states).  Such a ground-state can  suffer an OSMT when further increasing $U$ since the narrower band (more localized orbital) can become fully localized.  Next, adding the Hund's term (second choice) creates a preference for the two spin-aligned states in this four dimensional subspace by lowering their energy: this enhances inter-orbital correlations at the expense of intra-orbital correlations which favor an OSMT phase.  Third, adding the spin-flip and pair-hopping (third choice) terms essentially cancels the effect of the Hund's term.  This is explained by a straightforward computation of the matrix elements of this interaction in the four dimensional subspace.  One finds that the spin-flip term couples the two states where electrons have opposite spins with a strength that is precisely such that their symmetric combination has the same energy lowering as the Hund's term induces for the spin-aligned states.  Thus, we are essentially back to the four states we had when only operating with the intra-orbital interaction (first choice).  Our final comment is that these differences are not very dramatic once the hopping ratio $t_2/t_1$ is below $\approx0.5$.  As Fig.~\ref{fig:osmtphasehint} shows, in all cases only a modest value for $J$ is sufficient to stabilize the OSMT phase instead of a standard Mott transition.

\begin{figure}
\centering
\includegraphics[width=0.5\textwidth]{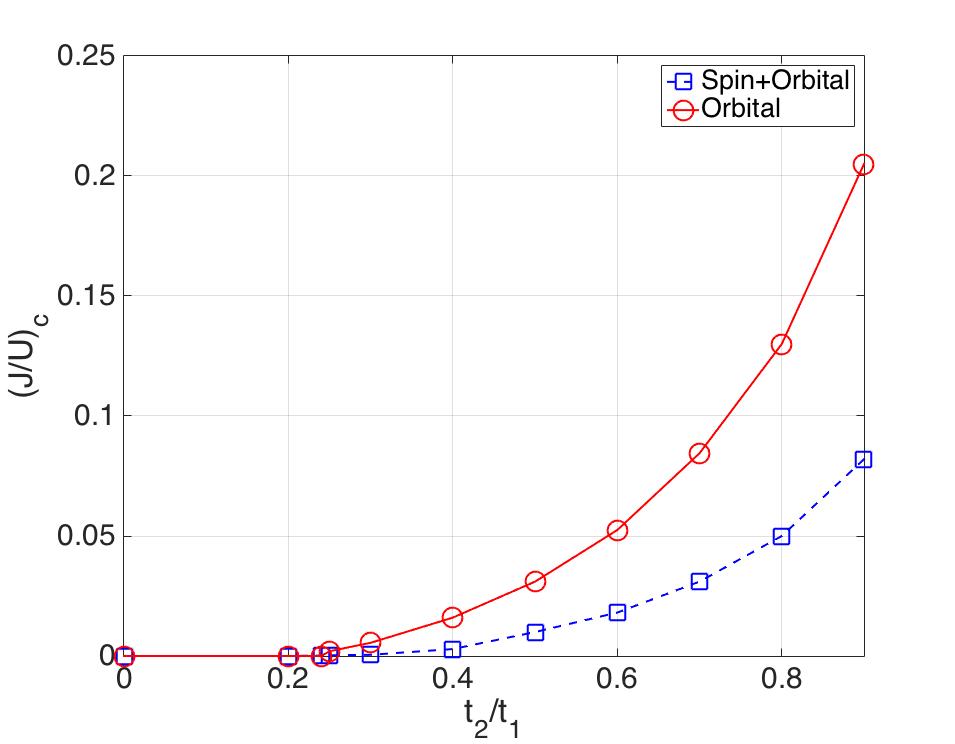}
\caption{Phase diagram for the anisotropic two-band single-site Hubbard model at half-filling as a function of the anisotropy ratio $t_2/t_1$ and $J$.  Two slave boson methods are used:  orbital slave (red circles) and spin+orbital slave (blue squares).  In each case, the boundary curve demarcates the possible existence of an Orbital-Selective Mott Transition when $U$ is ramped up from $U=0$.  Regions above the boundary display OSMT while regions below it present a standard Mott transition where both bands become insulating at the same critical $U_c$ value. }
\label{fig:osmtphase}
\end{figure}

\begin{figure}
\centering
\includegraphics[width=0.5\textwidth]{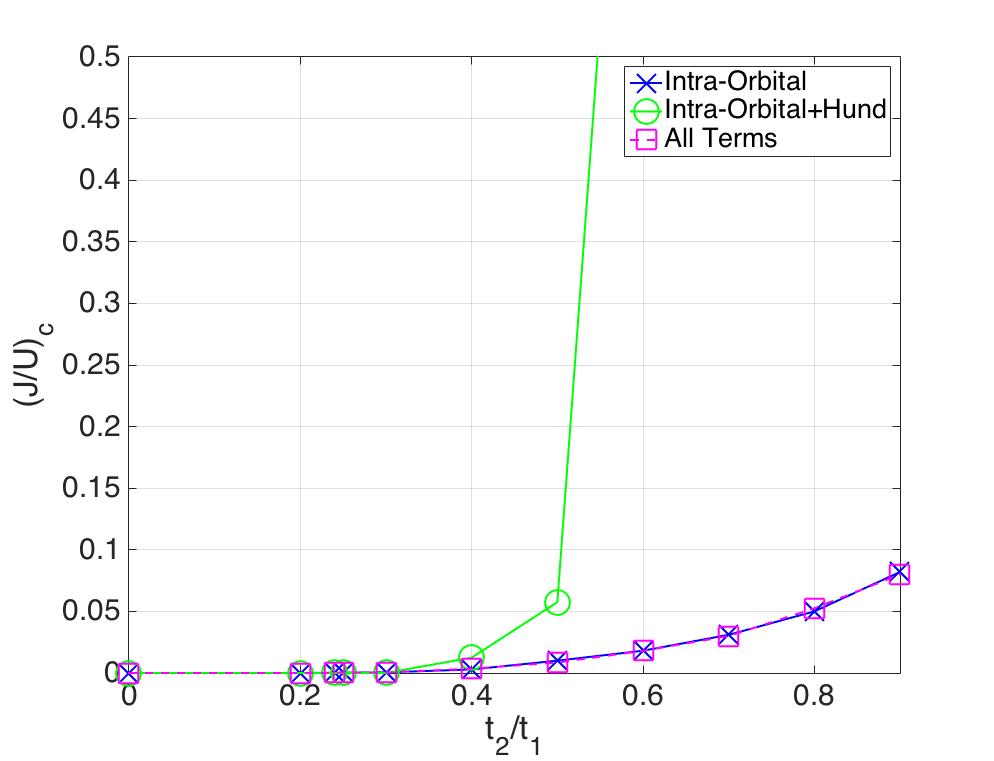}
\caption{Phase diagram for the anisotropic two-band single-site Hubbard model at half-filling as a function of the anisotropy ratio $t_2/t_1$ and $J$ for the spin+orbital slave model.  Three different  interaction terms are used: intra-orbital term only which is Eq.~(\ref{eq:hint2terms}), intra-orbital plus Hund's which is Eq.~(\ref{eq:Ham2}), and all terms included which is Eq.~(\ref{eq:Ham3}).}
\label{fig:osmtphasehint}
\end{figure}

\subsection{Ground State Energies}

A final and most stringent test for the slave models is to compare their total energies.  In the interest of space, we will focus on the simplest case of degenerate orbitals, isotropic hopping, and phases that are paramagnetic and paraorbital (no orbital differentiation) to make some general comments.   In a fully self-consistent model with more parameters and non-degenerate bands, we may expect more complexity to be revealed.  Previous work \cite{Quan2012} has shown that ground-state calculations can reveal competition between the orbital-selective Mott state (due to very large crystal-field splitting) and an anti-ferromagnetic Mott insulating state (due to a large $J$), a transition which is likely first-order\citep{Quan2012}. 

With $J=0$, the ground state energy per site of the paramagnetic and paraorbital phase is
\begin{equation}
E_g=-\sum_{\alpha}\sum_{m\in\alpha} t^{eff}_m \langle \hat O_{\alpha}\rangle_s +\frac{U}{2} \langle [\sum_{\alpha}\hat N_{\alpha}-n_{orb}]^2 \rangle \,.
\end{equation}
We compute the ground-state energy as a function of $U$ for one-band and two-band isotropic models at half-filling (same systems that are in the above sections) and also include the Hartree-Fock total energy.  Figures \ref{fig:onebandenergy} and \ref{fig:twobandenergy} display the energies versus $U$ for the one-band and two-band cases, respectively.  The plots employ the half-band width $D=2t$.

In all cases, for large enough $U$ the slave models produce an insulating phase (i.e., isolated atomic-like sites) which has zero hopping and zero number fluctuation and thus zero energy in this model.  The Hartree-Fock total energy necessarily has a linear dependence on $U$ for the high degree of spin and orbital symmetry since the Hartree-Fock Slater determinant wave function will be unchanged versus $U$ and always predicts a metallic system.  

The next observation is that for small $U$, some of the slave models do worse than Hartree-Fock.  However, as $U$ is increased their total energies eventually drop below the Hartree-Fock one.  Furthermore, increasing the number of bands from one to two improves the total energies of all slave methods compared to Hartree-Fock.  For a given number of bands, increasing the fine-grained of the slave model (i.e., having more slave modes per site) also lowers the total energy.  Hence, the slave-rotor is generally the worst performer.

A final observation is that only the fully fine-grained spin+orbital slave method always predicts a total energy below that of Hartree-Fock.  It also has the correct linear slope of $E_g$ versus $U$ matching the Hartree-Fock one.  The other slave methods have higher slopes of $E_g$ versus $U$ at the origin so that they can only outperform Hartree-Fock beyond some finite value of $U$.  The slope matching of the spin+orbital slave is a natural expression of its accounting in detail for all the quantum numbers on each site and in being forced (like all slave models) to reproduce the non-interacting state at $U=0$.  The fact that the other slave models have higher slopes is a reflection of their larger (and quantitatively incorrect) number fluctuations at $U=0$.  Namely, the interaction Hamiltonian $\hat H^{int}$ is a quadratic function of the occupancy numbers so that its expectation value (the interaction energy) depends directly on the fluctuations of these occupancies; at fixed $U$, the larger the set of allowed occupancies in a slave model, the larger this quadratic fluctuation and the higher the interaction energy.  In fact, the number fluctuations of the slave-rotor model are so large at $U=0$ that they lead to a pathological infinite slope of $E_g$ versus $U$ at $U=0$.  By comparison, the number slave method, which can be viewed as a corrected rotor, has a much more reasonable behavior.
\begin{figure}
\centering
\includegraphics[width=0.5\textwidth]{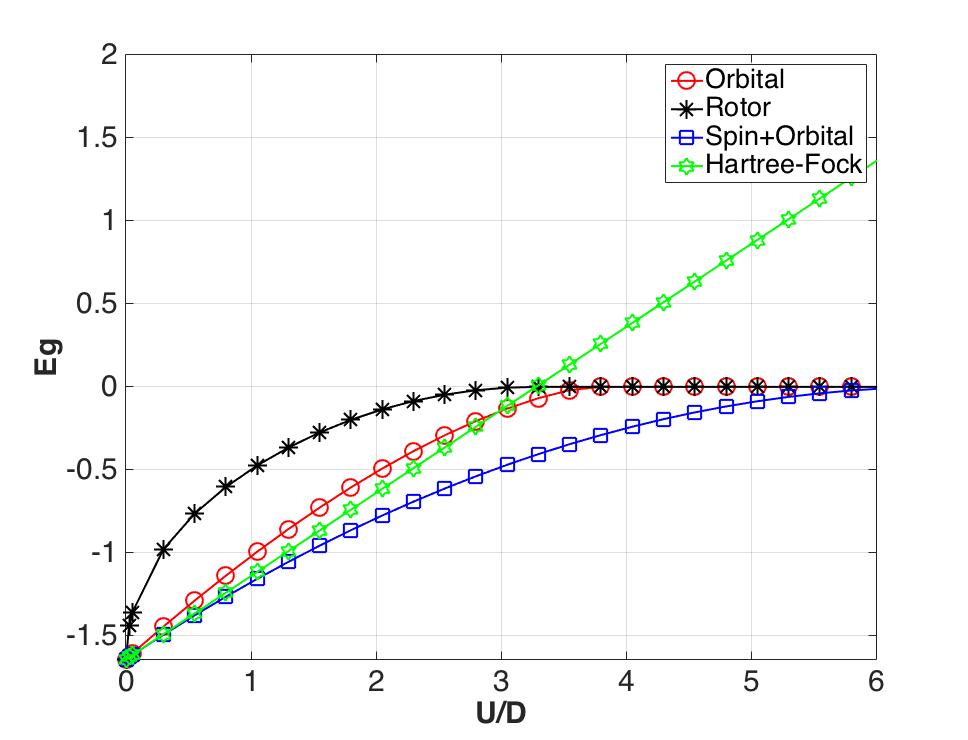}
\caption{Ground-state energy per site ($E_g/t$) of a single band Hubbard model at $J=0$ in the paramagnetic phase at half filling for a variety of slave representations as well as for the Hartree-Fock approximation.  $D=2t$ is the band width of the non-interacting system.  For this model the orbital slave is identical to the number slave and the spin slave is the same as the spin+orbital slave.}
\label{fig:onebandenergy}
\end{figure}

\begin{figure}
\centering
\includegraphics[width=0.5\textwidth]{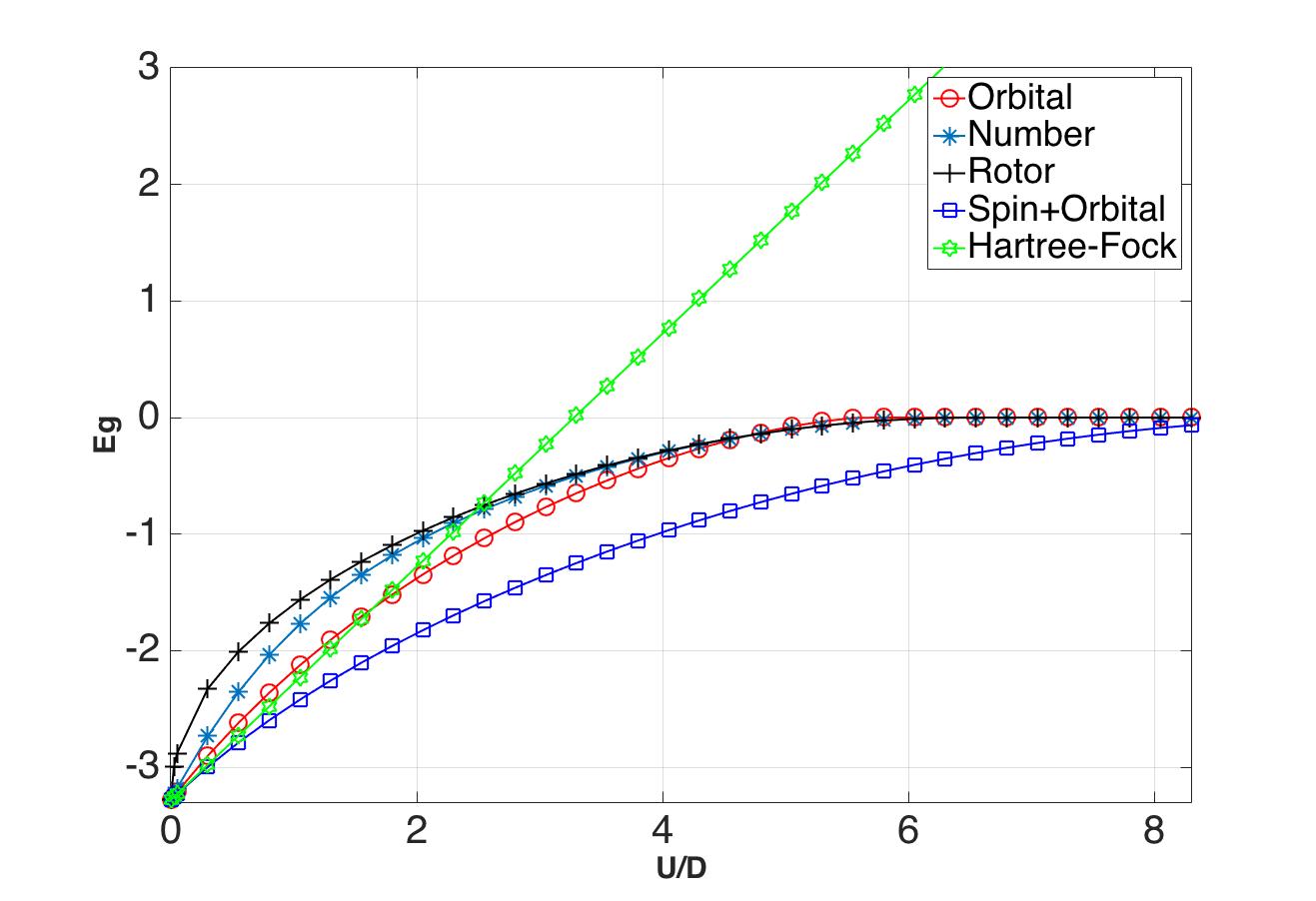}
\caption{Ground-state energy per site  ($E_g$) for an isotropic two-band Hubbard model at half filling for $J=0$ in the paramagnetic and paraorbital phase.}
\label{fig:twobandenergy}
\end{figure}

As a side note, it is interesting that for the single-band case, one has the following analytical results based on the coincidence of the of the spin+orbital slave and Gutzwiller approximations.  In the metallic phase,  where $U<U_c$, the quasiparticle weight $Z$ is given by
\begin{equation}
Z=1-U^2/U_c^2
\end{equation}
and from perturbation theory at small $Z$\cite{Florens2004}
\begin{equation}
U_c=8D\,.
\end{equation}
Using the following definition: 
\begin{equation}t_0=t\langle f^\dag_{im\sigma}f_{im\sigma}\rangle_{U=0}
\end{equation}
the ground-state energy is given by
\begin{equation}
E_g=-2t_0+\frac{U}{4}-\frac{U^2}{128t_0^2}\,.
\end{equation}
For the insulating state ($U\ge U_c$), we have $E_g=0$.

Our  calculations in this section permit us to say that while our generalized approach permit us to easily compare different slave models and isolate different degrees of freedom simply, total energy comparisons are much more challenging.  First, one should do energy comparisons of different phases within a single slave model since the differing models can produce differing total energies with dependence on the details of the system.  Second, after understanding the relevant degrees of freedom and how the influence the physical behavior, the total energy calculation should be most accurate with the most fine-grained model which is in the spin+orbital slave representation (``slave-spin'' in the literature).

\section{Conclusions}
\label{sec:conclusion}

We have developed a generalized formalism that reproduces previous slave-particle formalisms in appropriate limits but also allows us to define and explore intermediate models and to compare them systematically.  Our formalism moves beyond the analogy with angular momentum behind slave-rotor formalism, and instead works directly in the physically correct finite-sized number representation permitting new models to be developed in  a more natural way.  As an example, we have shown how the standard Mott transition as well as the orbital selective Mott transition appear in different slave models for single-band and two-band Hubbard models.

We believe it is useful to have a variety of slave particle methods on hand as they provide computationally inexpensive methods for exploring the role of electronic correlations in materials and interfaces with broken symmetries (e.g., orbital symmetry breaking).  The cheap computational load is particularly advantageous for interfacial systems where translational symmetry is lost in one direction and simulation cells that capture the region near the interface must contain at least tens to hundreds of atoms.  As such, these simpler slave-particle models are useful for exploratory research where more accurate and expensive Hubbard-model solvers such as DMFT\cite{Georges1996, Kotliar2006} would be prohibitive to apply routinely.  The ability to isolate potentially interesting correlated degrees of freedom from each other by choosing different slave approaches may illuminate which degrees of freedom are the most critical to model accurately.

\begin{acknowledgments} 
This work was supported  by the National Science Foundation via Grant MRSEC DMR 1119826. We thank Luca de’Medici for helpful discussions at the start of this work
\end{acknowledgments}

\appendix
\section{}
\label{appendix1}

In this appendix, we provide some detailed examples of how the physical subspace is isolated from the extended Hilbert space of spinon+slave boson states and how the operators act in the physical subspace.  In the process, we also provide explicit examples for various choices of the slave labels $\alpha$.  We focus on a single site $i$ and hence suppress the site label $i$ below.  

The original Hilbert space, i.e, the Fock space of the fermionic $\hat d_{m\sigma}$ field operators, is spanned by basis kets in the occupancy number representation for the field operators and have the form $\ket{\{n_{m\sigma}\}}$
where $n_{m\sigma}\in\{0,1\}$.  The enlarged Hilbert space for spinons and slave particles  is spanned by product kets in the number occupancy basis of the form
\[
\ket{\{n_{m\sigma}\}}_f\ket{\{N_{\alpha}\}}_s
\]
where, again, $n_{m\sigma}\in\{0,1\}$ are the fermionic spinon occupancies while $N_{\alpha}$ are the bosonic particle counts.  The $f$ and $s$ subscripts label the spinon and slave boson kets.

The constraint of Eq.~(\ref{eq:fmatchNconstraint}) on the physical allowed states translates to the numerical constraint
\begin{equation}
N_{\alpha} = \sum_{m\sigma\in\alpha} n_{m\sigma}\,.
\label{eq:numbermatch}
\end{equation}
We remember that we choose the $\{n_{m\sigma}\}$ to match exactly between the original electron and spinon kets.

We begin with the simplest example of a single spatial orbital on the site where the kets look like $\ket{n_\uparrow, n_\downarrow}_f\ket{\{N_\alpha\}}_s$.  There are two states for electrons and thus a total of four possible configurations: no electrons, one spin up electron, one spin down electron, and a pair of spin up and down electrons.  If we have a single slave boson per site to simply count the number of electrons so the $\alpha$ label is nil (i.e., the number slave representation), then our four physically allowed kets are
\[
\ket{0,0}_f\ket{0}_s \ , \ \ket{1,0}_f\ket{1}_s \ , \ \ket{0,1}_f\ket{1}_s \ , \ \ket{1,1}_f\ket{2}_s \,.
\]
We note that the number of slave particles is constrained by Eq.~(\ref{eq:fmatchNconstraint}) to the total the number of spinons.  

Next, if we have this single orbital but instead we choose to have a slave mode per spin channel (i.e, the spin+orbital slave representation), then we have two sets of slave bosons since now $\alpha=\sigma$.  The four physical states are now
\[
\ket{0,0}_f\ket{0,0}_s \ , \ \ket{1,0}_f\ket{1,0}_s \ , \ \ket{0,1}_f\ket{0,1}_s \ , \ \ket{1,1}_f\ket{1,1}_s \,.
\]

A more complex set of examples has two spatial orbitals per site.  Here we have four choices of spinon label $m\sigma$ which we order as 1$\uparrow$, 1$\downarrow$, 2$\uparrow$, 2$\downarrow$.  For the number slave representation, we have the 16 physical kets
\begin{multline*}
\ket{0,0,0,0}_f\ket{0}_s \ , \ \ket{1,0,0,0}_f\ket{1}_s \ , \ \ket{0,1,0,0}_f\ket{1}_s \ ,\\
\ket{0,0,1,0}_f\ket{1}_s  \ , \ \ket{0,0,0,1}_f\ket{1}_s  \ , \ \ket{1,1,0,0}_f\ket{2}_s  \ ,\\
 \ket{1,0,1,0}_f\ket{2}_s \ , \ \ket{1,0,0,1}_f\ket{2}_s\ , \ \ket{0,1,1,0}_f\ket{2}_s \ ,  \\
\ket{0,1,0,1}_f\ket{2}_s \ ,  \ \ket{0,0,1,1}_f\ket{2}_s \ ,  \ \ket{1,1,1,0}_f\ket{3}_s \ ,\\
 \ket{1,1,0,1}_f\ket{3}_s \ , \ \ket{1,0,1,1}_f\ket{3}_s \ , \ \ket{0,1,1,1}_f\ket{3}_s \ ,\\
 \ket{1,1,1,1}_f\ket{4}_s\,.
\end{multline*}
An orbital slave representation has slave bosons counting the number of electrons in each spatial orbital, so $\alpha=m$.  The 16 allowed kets are
\begin{multline*}
\ket{0,0,0,0}_f\ket{0,0}_s \ , \ \ket{1,0,0,0}_f\ket{1,0}_s \ , \ \ket{0,1,0,0}_f\ket{1,0}_s \ ,\\
\ket{0,0,1,0}_f\ket{0,1}_s  \ , \ \ket{0,0,0,1}_f\ket{0,1}_s  \ , \ \ket{1,1,0,0}_f\ket{2,0}_s  \ ,\\
 \ket{1,0,1,0}_f\ket{1,1}_s \ , \ \ket{1,0,0,1}_f\ket{1,1}_s\ , \ \ket{0,1,1,0}_f\ket{1,1}_s \ ,  \\
\ket{0,1,0,1}_f\ket{1,1}_s \ ,  \ \ket{0,0,1,1}_f\ket{0,2}_s \ ,  \ \ket{1,1,1,0}_f\ket{2,1}_s \ ,\\
 \ket{1,1,0,1}_f\ket{2,1}_s \ , \ \ket{1,0,1,1}_f\ket{1,2}_s \ , \ \ket{0,1,1,1}_f\ket{1,2}_s \ ,\\
 \ket{1,1,1,1}_f\ket{2,2}_s\,.
\end{multline*}
Alternatively, one can use the spin slave representation where the bosons count the number of electrons of each spin, so $\alpha=\sigma$.  The allowed kets are
\begin{multline*}
\ket{0,0,0,0}_f\ket{0,0}_s \ , \ \ket{1,0,0,0}_f\ket{1,0}_s \ , \ \ket{0,1,0,0}_f\ket{0,1}_s \ ,\\
\ket{0,0,1,0}_f\ket{1,0}_s  \ , \ \ket{0,0,0,1}_f\ket{0,1}_s  \ , \ \ket{1,1,0,0}_f\ket{1,1}_s  \ ,\\
 \ket{1,0,1,0}_f\ket{2,0}_s \ , \ \ket{1,0,0,1}_f\ket{1,1}_s\ , \ \ket{0,1,1,0}_f\ket{1,1}_s \ ,  \\
\ket{0,1,0,1}_f\ket{0,2}_s \ ,  \ \ket{0,0,1,1}_f\ket{1,1}_s \ ,  \ \ket{1,1,1,0}_f\ket{2,1}_s \ ,\\
 \ket{1,1,0,1}_f\ket{1,2}_s \ , \ \ket{1,0,1,1}_f\ket{2,1}_s \ , \ \ket{0,1,1,1}_f\ket{1,2}_s \ ,\\
 \ket{1,1,1,1}_f\ket{2,2}_s\,.
\end{multline*}

The final point is to check that the original electron operators $\hat d_{m\sigma}$ have the same effect as the combination of spinon and slave $\hat f_{m\sigma}\hat O_\alpha$ in the physical subspace.  That this is in fact true follows directly from the defining Equations (\ref{eq:dlowernumber}-\ref{eq:Olowernumber}) along with the constraint on $N_\alpha$ in Eq.~(\ref{eq:numbermatch}). It is easy to check that the matrix elements of $\hat d_{m\sigma}$ and $\hat f_{m\sigma}\hat O_\alpha$ must match:
\[
\bra{n'_{m\sigma}}\hat d_{m\sigma}\ket{n_{m\sigma}} = {}_f\bra{n'_{m\sigma}}\hat f_{m\sigma}\ket{n_{m\sigma}}_f\cdot{}_s\bra{N'_{\alpha}}\hat O_{\alpha}\ket{N_{\alpha}}_s\,.
\]
The matching of the $\hat d$ and $\hat f$ matrix elements is clear because the occupancies $n_{m\sigma}$ and $n'_{m\sigma}$ match by definition on both sides and both operators have identical behavior on the occupancies as per Eqs.~(\ref{eq:dlowernumber}) and (\ref{eq:flowernumber}).  Thus both sides are non-zero only if the $n'$ occupancy set has one fewer total count than the $n$ occupancy set.  As long as $N_\alpha>0$, the matrix element of $\hat O_\alpha$ is unity because $N'_\alpha = N_\alpha-1$ must be true due to the occupancy matching of Eq.~(\ref{eq:numbermatch}).  If $N_\alpha=0$, it must be that $n_{m\sigma}=0$, so that the matrix element of $\hat O_\alpha$ is irrelevant because the fermionic matrix elements (of $\hat d$ and $\hat f$) are both zero.

\bibliography{library}

\end{document}